\documentstyle[graphicx,amsmath,epsfig]{elsart}

\newcommand\ba{\begin{eqnarray}}
\newcommand\ea{\end{eqnarray}}
\newcommand\be{\begin{equation}}
\newcommand\ee{\end{equation}}
\newcommand\nn{\nonumber}

\begin{document}
\begin{frontmatter}
\title{Cross section of the processes $e^++e^-\to e^++e^-(\gamma)$,
$\to \pi^++\pi^-(\gamma)$, $ \mu^++\mu^-(\gamma)$, $ \gamma+\gamma(\gamma)$ in the energy region 200 MeV $\le 2E\le$ 3 GeV }
\author{E.~A.~Kuraev and A.~I.~Ahmadov}
\address{\it JINR-BLTP, 141980 Dubna, Moscow region, Russian Federation}
\author{S.~N.~Panov}
\address{\it Irkutsk State University, Irkutsk, Russian Federation}
\author{E.~Tomasi-Gustafsson}
\thanks{Corresponding author:etomasi@cea.fr}
\address{\it CEA,IRFU,SPHN, Centre de Saclay, F-91191 Gif-sur-Yvette, France}

\begin{abstract}
The cross section for different processes induced by $e^+e^-$ annihilation, in the kinematical
limit $\beta_{\mu}\approx\beta_{\pi}=\left(1-m_{\pi}^2/\epsilon^2\right )^{1/2}\sim 1$,
is calculated taking into account first order corrections to the amplitudes and the corrections
due to soft emitted photons, with energy $\omega\le\Delta E\le \epsilon$ in the center of mass
of the $e^+e^-$ colliding beams. The results are given separately for charge--odd and charge--even terms in the final channels $\pi^+\pi^-(\gamma)$ and $\mu^+\mu^-(\gamma)$. In case of pions,
form factors are taken into account. The differential cross sections for the processes:
$e^++e^-\to e^++e^-(+\gamma)$, $\to \pi^++\pi^-(\gamma)$, $\to \mu^++\mu^-(\gamma),\to \gamma\gamma(\gamma)$
have been calculated and the corresponding formula are given in the ultrarelativistic limit
$\sqrt{s}/2= \epsilon \gg m_{\mu}\sim m_{\pi}$ . For a quantitative evaluation of the contribution of higher
order of the perturbation theory, the production of $\pi^+\pi^-$, including radiative corrections,
is calculated in the approach of the lepton structure functions. This allows to estimate the precision of the obtained results as better than 0.5\% outside the energy region corresponding to narrow resonances.
A method to integrate the cross section, avoiding the difficulties which arise from singularities is also described.
\end{abstract}
\end{frontmatter}
\maketitle
\section{Introduction}

The measurement of the cross section for hadronic or leptonic final states in the experiments
on colliding $e^+e^-$ beams, with a precision of 1 or 2\% is a current problem, at different
world accelerators:  Frascati, Novosibirk, Bejing.. This work aims to  calculate the cross
section of different processes with a precision of this level. This paper is composed by six 
Sections followed by Conclusions.

In the first part we calculate first order corrections to the amplitudes for the process of creation of a pion pair: $e^++e^-\to \pi^++\pi^-(+\gamma)$
we consider the emission of soft photons and derive the differential cross section as a function of
the energy, of the angles of the photons with respect to the direction of the beam:
$\theta=\widehat{\vec q_-\cdot\vec p_-}$ ( $\vec q_- $ ($\vec p_-$) is the three momentum of the photon
(beam), and of the parameter $\Delta\epsilon/\epsilon$, ($\Delta\epsilon$ is the maximum energy
of the soft photons which escape the detection,$ \epsilon$ is the beam energy). Charge--odd and charge--even contributions of the corrections to the cross section are calculated. Their asymptotic
expressions are also given and shown that they are in agreement with a result previously found. For
the calculation of the contribution of the box-diagram,  we make the assumption that the pions are
without structure. We also consider  the cross section of pion pair creation, with emission of an
additional hard photon, taking into account the form factors of pions. We calculate the cross section
in two cases: for $\beta_{\pi}\sim 1$ and in the ultrarelativistic case $1-\beta_{\pi}\ll 1$.

In the second part, we calculate in a similar way the process  $e^++e^-\to \mu^++\mu^-(\gamma)$.
The contributions of virtual and soft photons in the charge-odd and in 
charge--even parts of the
cross section is given for $\beta\sim 1$ and for $\beta \to 1$.  The differential cross section of
this process is also derived.

The third part is devoted to the process $e^++e^-\to e^++e^-(\gamma)$ with emission of virtual and
soft photons and to the calculation of the cross section of this process when additional hard photon
are emitted in ultrarelativistic kinematics. The cross section of the process  $e^++e^-$ with
subsequent emission of any number of emitted photons
(virtual, soft, and hard as well), is calculated in the large logarithm (LLA) approximation.
We express the cross section in form of a Drell-Yan process \cite{BCM83}:
\be
d\sigma=\int dx_1\int dx_2 {\cal D}(x_1,\beta_q) {\cal D}(x_2,\beta_q) d \tilde\sigma_0
{\cal D}\left( \frac{y}{y_1},\beta_q\right ) {\cal D}\left( \frac{z}{z_1},\beta_q\right )\frac{1}{y_1z_1},
\label{eq:eq1}
\ee
where ${\cal D}(x,\beta)$ is the lepton structure functions  (LSF) which describes the probability to find an electron (positron) of a given energy and four-momentum squared $q^2$ in the initial electron (positron). For the emission at large angles one obtains:
$$
\beta_q\sim \beta\frac{2\alpha}{\pi}(\rho_e-1),
~\rho_e=\ln\frac{s}{m_e^2},~s=4\epsilon^2,
~d\tilde\sigma_0\sim\left( 1-\frac{\alpha}{3\pi}\rho_e\right )^{-2}d\hat\sigma_0,
$$
where $d\hat\sigma_0$ is the Born cross section for the scattering $e^++e^-\to e^{+ '} + e^{-'}$.

In sections IV and V we derive the formulas similar to Eq. (\ref{eq:eq1}), which hold for the processes  $e^++e^-\to \mu^+ + \mu^-(\gamma)$ and $e^+ + e^-(\gamma)$. We discuss the expression of the cross section, and compare it with the exact result at first order of the perturbation theory as well as with the contribution of higher order terms in the large logarithm approximation:
\be
\frac{\alpha}{\pi}\rho_e=1.
\label{eq:eq2}
\ee
The evaluation of the error on the term:
\be
\left( \frac{\alpha}{\pi}\right )^2\rho_e\sim 0.01\%
\label{eq:eq3}
\ee
defines the precision obtained in this model (which is largely sufficient for most practical application).

If the calculation is limited  only to the corrections at the lowest order in PT (as obtained in Sections 1-3), then the precision is determined by the next terms of this approach, which are of the order of:
\be
\left (\frac{\alpha \rho_e}{\pi}\right )^2\sim\frac{1}{400} \sim 0.2\%,
\label{eq:eq4}
\ee
which is sufficient for the comparison of the results of the current experiments (except in the region of narrow resonances, where formula (\ref{eq:eq1}) does not apply).

In Section VI we discuss the annihilation process of electron--positron in two (three) gammas, and give the cross section in frame of the LSF method.

In Conclusions we discuss and develop a method for integrating the cross section, based on the method of quasi-real electrons \cite{Ba73}, in the kinematical region where the photon is emitted at small angle from one of the charged particles.
 \section{ The process  $e^++e^-\to \pi^+ + \pi^-(\gamma ) $}
\subsection{  First order correction to the Born amplitude}
\begin{figure}
\begin{center}
\includegraphics[width=15cm]{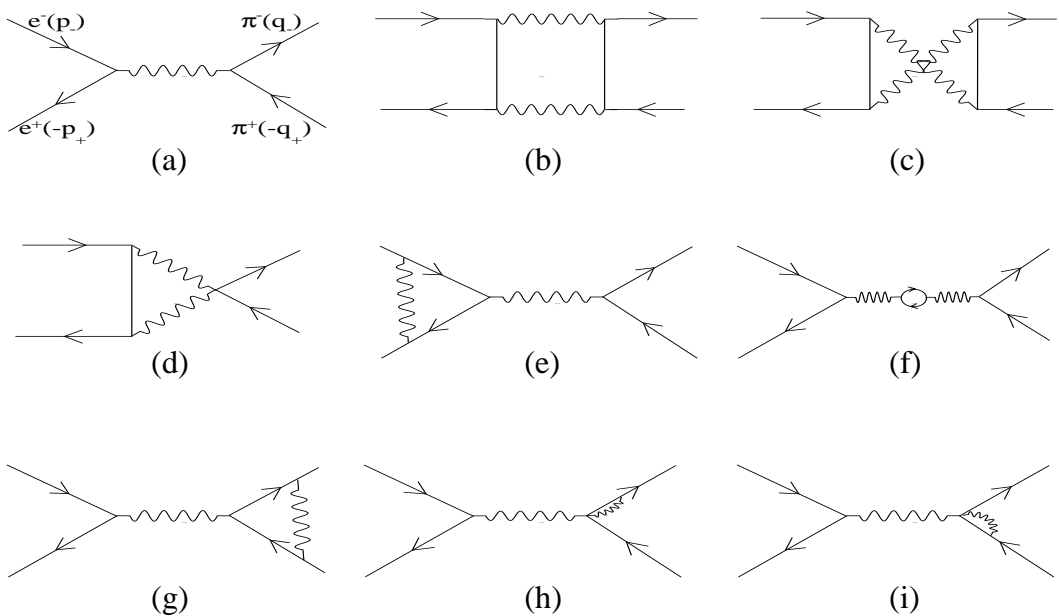}
\caption{\label{Fig:fig1}Feynman Diagrams for the process $e^++e^-\to  \pi^++\pi^-$.}
\end{center}
\end{figure}

The corresponding Feynman diagrams are shown in Fig. \ref{Fig:fig1}.
In order to calculate the cross section, it is necessary to calculate the moduli squared of the amplitudes, summed over the spin states.

The contribution of the diagram of Fig. \ref{Fig:fig1}a to the cross section is:
\be
\frac{d\sigma_0}{d\Omega}=\displaystyle\frac{\alpha^2\beta^2}{s}\sin^2\theta\left |F_{\pi}(s)\right |^2, 
\label{eq:eq11}
\ee
$$\theta=\widehat{\vec q_-\cdot\vec p_-},~\beta=\left(1-\frac{m_{\pi}^2}{\epsilon^2}\right )^{1/2},~ s=4\epsilon^2,~m=m_{\pi},$$
where $F_{\pi}(s)$ is the pion form factor. We show below that, neglecting the terms of fourth order, the contributions of the squares of the one--loop diagrams (Fig. \ref{Fig:fig1}b-i) to the cross section cancel. Such contributions can be taken into account at higher order with the method of LSF, that will be discussed in Section III.

Let us start from the charge-odd part of the cross section.
The matrix element, corresponding to the diagrams Fig. \ref{Fig:fig1}a-d has the form:
\ba
{\cal M}&=& -\frac{8\pi\alpha i}{s}\overline{v}\hat Q u +i\alpha^2 \overline{v}\int \frac{d^4k}{i\pi^2}
\left [ \displaystyle\frac{(\hat k-\hat Q-\hat q_+)(\hat k-\hat \Delta)(\hat k-\hat Q+\hat q_-)}{(+)(-)(\Delta)(Q)}
+ \right .
\nn\\
&&\left .\displaystyle\frac{(\hat k+\hat Q-\hat q_-)(\hat k-\hat \Delta)(hat k+\hat Q+\hat q_+)}{(+)(-)(\Delta)(\tilde Q)}
-\displaystyle\frac{2\gamma^{\lambda}(\hat k-\hat \Delta)\gamma_{\lambda}}{(+)(-)(\Delta)}
\right ]u.
\label{eq:eq12}
\ea
with the following notations:
\ba
\Delta&=&\frac{1}{2}(p_+-p_-),~p=\frac{1}{2}(p_++p_-)=\frac{1}{2}(q_++q_-),\nn\\
Q&=&\frac{1}{2}(q_+-q_-),~s=(p_++p_-)^2=4\epsilon^2,~q^2_+=q^2_-=m^2,~p^2_+=p^2_-=0,\nn\\
(\pm )&=&(k\mp p)^2-\lambda^2,~(Q)=(k-Q)^2-m^2,~(\tilde Q)=(k+Q)^2-m^2,\nn\\
(\Delta)&=& (k-\Delta)^2-m_e^2,~\tau=m^2-t,~t=(p_--q_-)^2,~u=(p_--q_+)^2.
\label{eq:eq13}
\ea
$\overline{v}$, $u$ are the spinors which represent the initial positron and electron:
\be
\overline{v}\hat{p}_+=\hat{p}_- u=0.
\label{eq:eq14}
\ee
Performing the loop-integral over the four-momentum, the following integrals,  which coincide with the corresponding
ones for the process $e^++e^-\to \mu^+ + \mu^-$, need to be calculated:
\ba
&&\int\displaystyle\frac{d^4k}{i\pi^2}\displaystyle\frac{\{1;k^{\mu}\}}{(+)(-)(Q)(\Delta)}=\{J;J_{\Delta}\Delta^{\mu}+J_QQ^{\mu}\},\nn\\
&&\int\frac{d^4k}{i\pi^2}\displaystyle\frac{\{1;k^{\mu}\}}{(Q)(\Delta)(\pm)}=\{H;\pm H_{\rho}\rho^{\mu}+H_{\Delta} \Delta^{\mu}+H_Q Q^{\mu} \},\nn\\
&&\int\frac{d^4k}{i\pi^2}\displaystyle\frac{\{1;k^{\mu}\}}{(Q)(+)(-)}=\{F_Q;Q^{\mu}G_Q\},
\int\frac{d^4k}{i\pi^2}\displaystyle\frac{
\{1;k^{\mu}\}}{(\Delta)(+)(-)}=\{F_{\Delta};G_{\Delta} \Delta^{\mu} \}.
\label{eq:eq15}
\ea
where:
\ba
J&=&-\displaystyle\frac{2}{s\tau}\ln \frac{s}{\lambda^2}\ln\frac{\tau}{mm_e},
~F_{\Delta}= \displaystyle\frac{1}{s}
\left( \frac{\pi ^2}{6}+\frac{1}{2}\rho_e^2 \right ),\nn\\
F_Q&=&\displaystyle\frac{1}{s\beta} \left [ \frac{\pi^ 2}{6}+\frac{1}{2}\rho^2+
2\rho\ln  \frac{1+\beta}{2} -4Li_2\left( \frac{1-\beta}{2} \right ) -2Li_2\left( -\frac{1-\beta}{1+\beta}\right )\right ],\nn\\
 H&=&-\displaystyle\frac{1}{2\tau}\left [ 2\ln\frac{\tau}{mm_e}\ln\frac{m^2}{\lambda ^2} +
 \ln^2\frac{\tau}{m^2} - \frac{1}{2}\ln^2\frac{m^2}{m_e^2}-2Li_2\left(-\frac{t}{\tau}\right )\right ],
\nn\\
J_{\Delta}&=&\displaystyle\frac{1}{2d}[(\Delta Q-Q^2)F+\Delta QF_{\Delta}-Q^2F_Q],\nn\\
J_Q&=&\displaystyle\frac{1}{2d}[(\Delta Q-\Delta^2)F+\Delta Q F_Q-\Delta ^2F_\Delta],\nn\\
G_{\Delta}&=&\displaystyle\frac{1}{s}\left (-2\rho_e+sF_\Delta\right ),~
G_Q=\displaystyle\frac{1}{s\beta^2}\left (-2\rho+sF_Q\right ), H_\rho=H+\displaystyle\frac{2}{\tau}\ln \frac{\tau}{mm_e},\nn\\
H_Q&=&\displaystyle\frac{1}{t}\ln \frac{\tau}{m^2},~
H_{\Delta}=-\displaystyle\frac{1}{\tau}\left [ \frac{\tau}{t} \ln \frac{\tau}{m^2}+2\ln \frac{\tau}{mm_e}\right],\nn\\
F&=&\displaystyle\frac{1}{2}sJ-H,~d=\Delta^2Q^2-(\Delta Q)^2,~\rho=\ln\frac{s}{m^2},~\rho_e=\ln\frac{s}{m_e^2},\nn\\
&&Li_2(z)=-\int_0^z\frac{dx}{x}\ln(1-x).\nn
\ea
First of all, let us note that in Eq. (\ref{eq:eq12}) the contribution of the last term of the sum in the square parenthesis is of the order of $\sim m_e/m_{\pi}\to 0$, which follows from the form of the corresponding integrals (\ref{eq:eq14}), (\ref{eq:eq15}). The contribution of the second term in Eq. (\ref{eq:eq12}), can be obtained from the first term, with the help of the substitution: $t\leftrightarrow u$ and changing all signs. As a result of straightforward, but lengthy calculations, we find the following expression for the modulus squared of the matrix element, summed over the spin states:
\ba
\sum_s|{\cal M}|^2&= &32 \pi^2\alpha^2\beta^2(1-\cos^2\theta)+32\pi \alpha^3\frac{1}{s}Re[-4d(-8\Delta QJ+\nn\\
&&G_Q-4H)+(F+F_{\Delta})(-16(\Delta Q)^2+16(\Delta ^2(\Delta Q)+\nn\\
&&8\Delta ^2 Q^2 -8\Delta ^4)+(F+F_Q)\Delta Q(8Q^2-16\Delta Q+8\Delta^2)-
\nn\\
&&
(Q\leftrightarrow -Q)].
\label{eq:eq16}
\ea
The coefficient of charge asymmetry, $\eta$ can be built from Eq. (\ref{eq:eq16}) as the ratio of differential cross sections at corresponding angles:
\be
\eta=\displaystyle\frac{
\displaystyle\frac{d\sigma}{d\Omega} (c)-
\displaystyle\frac{d\sigma}{d\Omega} (-c)}
{
\displaystyle\frac{d\sigma}{d\Omega} (c)+
\displaystyle\frac{d\sigma}{d\Omega} (-c)
},
~c=\cos\theta=\cos(\widehat{\vec q_-,\vec p_-}).
\label{eq:eq17}
\ee
The denominator can be calculated if the Born cross section is known (\ref{eq:eq11}). The one loop correction to the variable $\eta$ , after simplifying Eq. (\ref{eq:eq16}), and with the help of Eq. (\ref{eq:eq15}), is:
\ba
\eta^{virt}&=&\frac{\alpha}{\pi}
\left \{ 2\ln \displaystyle\frac{2\epsilon}{\lambda}\ln\displaystyle\frac{1+\beta c}{1-\beta c}
+\displaystyle\frac{1}{\beta^2\sin^2\theta}
\right . 
\left \{ (1-\beta c)\left [-l^2_-+2\rho L_- + 2\ell_-L_- -\nn \right . \right .\\
&&
2Li_2\left (\displaystyle\frac{1-\beta^2 }{2(1-\beta c)} \right )
- \displaystyle\frac{(1-\beta)^2}{2\beta} \left ( \frac{1}{2}\rho^2+\frac{\pi^2}{ 6}\right )
 +\nn\\
&&
\left . +\displaystyle\frac{1+\beta^2}{\beta}\left( \rho\ln \displaystyle\frac{2}{1+\beta}
-Li_2\left( -\displaystyle\frac{1-\beta }{1+\beta}\right ) +2Li_2\left( \displaystyle\frac{1-\beta }{2}\right )\right )\right ]+\nn \\
&&
\left .(1-\beta^2)\left [ \frac{1}{2} l^2_- -L_-(\rho+\ell_-)=LI-2\left(
\displaystyle\frac{1-\beta^2 }{2(1-\beta c) }\right ) \right ]\right \}\nn \\
&&
\left .
-(c\leftrightarrow -c)\right \},
\label{eq:eq18}
\ea
where
$$L_-=\ln\left [ 1- \displaystyle\frac{1-\beta^2 }{2(1-\beta c) }\right ];~
\ell_-=\ln \displaystyle\frac{(1-\beta c)}{2 }.
$$
In the ultrarelativistic limit, $1-\beta^2=m^2/\epsilon^2\ll 1$ :
\be
\eta^{virt}_{as}= \displaystyle\frac{\alpha }{\pi}\left [ -8\ln \left ( \displaystyle\frac{2\epsilon}{\lambda}\right )\ln\tan\displaystyle\frac{\theta}{2} -
2\displaystyle\frac{\ln^2\sin \displaystyle\frac{\theta}{2}}{\cos^2 \displaystyle\frac{\theta}{2}}
+2\displaystyle\frac{\ln^2\cos \displaystyle\frac{\theta}{2}}{\sin^2 \displaystyle\frac{\theta}{2}}
\right ].
\label{eq:eq18a}
\ee
Let us consider now the diagrams Fig. \ref{Fig:fig1}d,i. Their contribution to the cross section is an even function of $\cos\theta$ and it is written as:
\be
\displaystyle\frac{d\sigma ^{virt}_{even}}{d\Omega}=
2\displaystyle\frac{d\sigma_0}{d\Omega} \left( Re F_1^{(e)} + Re F^{(\pi)} - Re \Pi \right ).
\label{eq:eq19}
\ee
The quantities
 \ba
 Re F_1^{(e)}&=&\displaystyle\frac{\alpha }{\pi}\left [\left( \frac{m_e}{\lambda} -1\right ) (1-\rho_e) - \displaystyle\frac{1}{4}\rho_e^2+\frac{\pi^2}{3}- \displaystyle\frac{1}{4}\rho_e \right ],~
 \nn\\
 Re F^{(\pi)} &=&\frac{\alpha}{\pi}\left [ \frac{1}{2} \beta ^{\mu}
 \left (1-\frac{1}{3}\beta_{\mu}^2\right)\ln \frac{1-\beta^{\mu}}{1+\beta^{\mu}}- \frac{8}{9}+
 \frac{1}{3}\beta^2_{\mu}\right ]=\frac{\alpha}{\pi}\Delta_{\mu},\nn\\
 - Re \Pi&=&\frac{\alpha}{\pi}\left (\frac{1}{3}\rho_e-\frac{5}{9}\right ) +\delta_{\mu}+\delta_H,~\beta_{\mu}=\left( 1-\frac{m_{\mu}^2}{\epsilon^2}\right )^{1/2},
\label{eq:eq110}
\ea
are known expressions \cite{AB81} for the corrections to the Dirac part of the electron form factor to the vacuum polarization, due to electrons and muons. The quantity $\delta_H$ represents the contribution of the vacuum polarization:
\be
\delta_H=-\frac{s}{4\pi^2\alpha}{\cal P}\int_{4m_{\pi}^2}^{\infty}\frac{ds'\sigma_h(s')}{s'-s}
\sim\frac{\alpha}{\pi}\Delta_H, 
 \label{eq:eq111}
\ee
where the symbol ${\cal P}$ means that the integral is taken in principal sense;
$\sigma_h(s)$ is the cross section for hadron production in $e^+e^-$ collisions.
If one refers only to the channel $e^++e^-\to \pi ^++\pi^-$, then Eq. (\ref{eq:eq111}) can be approximated to:
\be
\sigma_h(s)\to \frac{pi\alpha^2}{3s}\beta_{\pi}^3;~\delta_H\to \delta_{\pi}= \frac{\alpha}{\pi}
\left (\frac{1}{12}\ln \frac{1+\beta_{\pi}}{1-\beta_{\pi}}-\frac{2}{3}-2\beta^2_{\pi}\right )\equiv
\frac{\alpha}{\pi}\Delta_{\pi}.
\label{eq:eq111a}
\ee
Finally, let us consider the diagrams in Fig. \ref{Fig:fig1}g-i,, which contain the corrections to the LSF of the pion:
\be
F^{\mu}(q_-,q_+)=(q_--q_+)^{\mu}e\left [1+F^{(2)}(q^2)\right ],~q=q_--q_+,~F^{(2)}(0)=0.
\label{eq:eq111b}
\ee
The standard procedure consists in joining the denominators of the Feynman diagrams and in integrating over the four-momentum loop. After regularization, one obtains:
\ba
F^{(2)}(q^2)&=&\frac{\alpha}{4\pi}
\left \{ (q^2-2m^2)\int_0^1dx\frac { \ln m^2/\lambda^2-2+\ln \left 
[(1-q^2/m^2)x(1-x)\right ]}{m^2-q^2x(1-x)}\right .
\nn\\
&&
 \left .
 + 2\left ( \ln\frac{m^2}{\lambda^2}-2\right ) \right \}=
\nn \\
 &=&\frac{\alpha}{4\pi}
 \left \{ 2\left ( \ln\frac{m^2}{\lambda^2}-2\right )+\frac{1+\beta^2}{\beta}
 \left [ \left ( \ln\frac{q^2}{\lambda^2}-2\right ) \ln \frac{1-\beta}{1+\beta}-
 \right .\right .\nn\\
 &&
 \frac{1}{2}\ln^2  \left (\frac{1+\beta}{2}\right )+
 \frac{1}{2} \ln^2  \frac{1-\beta}{2} +\ln\beta \ln \frac{1-\beta}{1+\beta}
 -\nn\\
 &&
  \left .\left .
  Li_2 \left(-\frac{1-\beta}{2\beta}\right )+Li_2\left(\frac{1+\beta}{2\beta}\right )\right]\right \}.
\label{eq:eq112}
\ea
Inserting this expression, Eq. (\ref{eq:eq19}) takes the form:
\ba 
\left(\frac{d\sigma}{d\Omega}\right )^{even}_{virt}&=&\frac{d\sigma_0}{d\Omega}\displaystyle\frac{2\alpha}{\pi}
 \left \{\left( \ln\frac{m_e}{\lambda }-1\right ) (1-\rho_e)
 -\frac{1}{4}\rho_e^2+ \frac{\pi^2}{3}
 -\frac{1}{4}\rho_e+ \right .
 \nn\\
 &&
 \left (\ln\frac{m}{\lambda }-1\right ) \left (1- \displaystyle\frac{1+\beta^2}{2\beta}\ln \frac{1+\beta}{1-\beta}\right )
 -  \frac{1+\beta^2}{8\beta} \left( \rho^2- \frac{\pi^2}{3}\right )
 + 
  \nn\\
 &&
 \frac{1}{3}\rho_e-\frac{5}{9}+\frac{1+\beta^2}{2\beta}\left [ -\rho\ln\beta +\frac{1}{2}\ln^2  \left (\frac{1+\beta}{2}
 \right )
 -\frac{1}{2}\ln^2\beta -\right .
 \nn\\
 &&
 \left .
 \ln\beta \ln \frac{1+\beta}{2}
 -Li_2\left(-\frac{1-\beta}{1+\beta}\right )\right] +
  \Delta_{\mu}+\Delta_H \Big \}.
 \label{eq:eq113}
\ea

 \section{Emission  of an additional photon}

\begin{figure}
\begin{center}
\includegraphics[width=15cm]{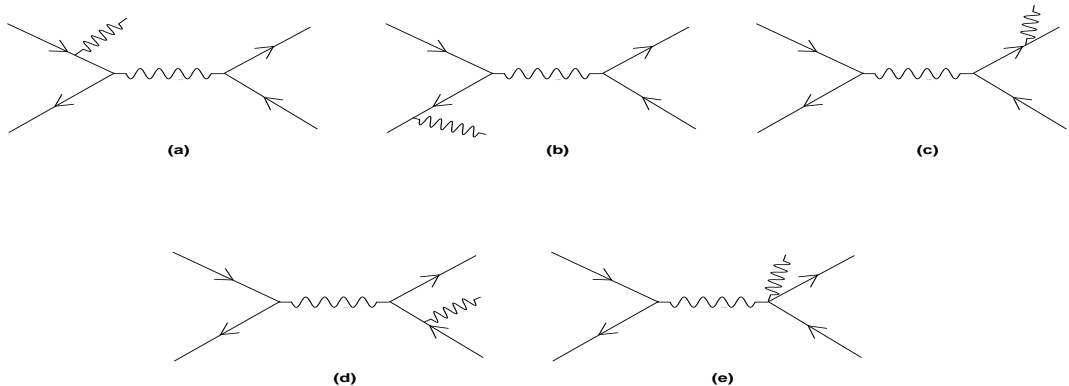}
\caption{\label{Fig:fig2}Feynman Diagrams for the process $e^++e^-\to  \pi^++\pi^-(\gamma)$.}
\end{center}
\end{figure}

 The Feynman diagrams corresponding to the process
 \be
 e^+(p_+)+e^-(p_-)\to \pi^+(q_+)+\pi^-(q_-)+\gamma(k),
 \label{eq:eq114}
\ee
in the Born approximation, are shown in Fig. \ref{Fig:fig2}. Let us start from the case of a soft photon in CMS:
\be
 k_0=\omega<\Delta\epsilon\ll\epsilon.
 \label{eq:eq115}
\ee
The modulus squared of the amplitudes of the process
(\ref{eq:eq114}), summed over the spin states of all particles, $\sum|{\cal M}|^2$, differs from the corresponding quantity for the process  $e^++e^-\to \pi^++\pi^-$ in the Born approximation, $\sum|{\cal M}_0|^2$, due to the emission of the accompanying photon:
\be
  \sum|{\cal M}|^2=-4\pi\alpha\left ( \displaystyle\frac{p_-}{p_-k} -\displaystyle\frac{p_+}{p_+k}
 + \displaystyle\frac{q_+}{q_+k}  -\displaystyle\frac{q_-}{q_-k}\right )^2
 \sum|{\cal M}_0|^2
  \label{eq:eq116}
\ee
Let us consider first, the part that changes sign when interchanging the four-momenta of the mesons:
$$
 \displaystyle\frac{p_+q_-}{p_+k\cdot q_-k} -
 \displaystyle\frac{p_-q_-}{p_+k\cdot q_-k} -
 \displaystyle\frac{p_+q_+}{p_+k\cdot q_+k} +
 \displaystyle\frac{p_-q_+}{p_-k\cdot q_+k}.
$$
Its contribution to the coefficient of charge asymmetry:
$$
\eta ^{soft}=4\left  (-\displaystyle\frac{4\alpha\pi}{16\pi^3}\right )
\int_{\epsilon<\Delta\epsilon}
\displaystyle\frac{d^3k}{\omega\cdot q_-k}
\left ( \displaystyle\frac{p_+q_-}{p_+k}- \displaystyle\frac{p_-q_-}{p_-k}\right )
$$
coincides with the corresponding expression, given below, for
 the process $e^++e^-\to \mu^++\mu^- (\gamma)$, after replacing $m_{\pi}\to m_{\mu}$.
Let us give the result for the total contribution $\eta=\eta^{soft}+\eta^{virt}$:
\ba
\eta&=&
 \displaystyle\frac{2\alpha}{\pi}
 \left \{ \ln \frac{1-\beta c}{1+\beta c }
 \ln\frac{\Delta\epsilon}{\epsilon}+\frac{1}{2}l^2_-
 -Li_2\left(-\frac{\beta^2(1-c^2)}{1+\beta^2+2\beta c}\right )-
 \frac{(1-\beta c)l^2_-}{2\beta^2(1-c^2)} + \right .\nn\\
 &&
 \left [-Li_e\left(\frac{1-\beta^2}{2(1-\beta c)}\right )+(\rho +\ell_-)L_-\right ]
 \left [-1+ \frac{1+\beta^2-2\beta c}{2\beta^2(1- c^2)}\right ]+\nn\\
 &&
 \frac{c}{4\beta^2(1- c^2)}
 \left [ (1-\beta )^2 \left (\frac{1}{2}\rho^2+ \frac{\pi^2}{6 }\right )
 -2(1+\beta )^2 \left (\rho\ln \frac{2}{1+\beta} - \right . \right .\nn\\
 &&
 \left . \left . Li_2 \left(-\frac{1-\beta}{1+\beta}\right )
 +2Li_2 \left(\frac{1-\beta}{2}\right) \right )\right]
 + \frac{(1-\beta^2)l^2_-}{4\beta^2(1- c^2)}
 +  \nn\\
 &&
 \int_0^{1-\beta^2}\frac{dx}{x}\left ( 1-\frac{x}{\chi_t}\right ) ^{-1/2} f(x) - (c\leftrightarrow -c ) \Big \},
 \label{eq:eq117}
\ea
 with:
 \ba \chi_t&= &\frac{(1-\beta c)^2}{ 1+\beta^2-2\beta c},
 \nn\\
 f(z)&=&\frac{1}{2}\left (\frac{1}{\sqrt{1-z}} -1\right ) \ln\frac{z}{4}
 - \frac{1}{\sqrt{1-z}}\ln\frac{\sqrt{1-z} +1}{2}, \nn
 \label{eq:eq117a}
\ea
and the other notations were given in (\ref{eq:eq18}).
In the ultrarelativistic limit, this result coincides with the one previously found in \cite{Be73,Ku77}:
 \ba
 \eta_{as}&=& \displaystyle\frac{2\alpha }{\pi}\left\{ 4\ln \tan\frac{\theta}{2}\ln \frac{\Delta\epsilon}{\epsilon} + \left (2- \frac{1}{\cos^2(\theta/2)}\right ) \ln^2\sin \frac{\theta}{2} -\right .
 \nn\\
 &&
 \left .\left( 2-\frac{1}{\sin^2(\theta/2)}\right ) \ln^2\cos\frac{\theta}{2} +
 \int ^{\sin^2(\theta/2)}_{\cos^2(\theta/2)}\frac{dx}{x}\ln(1-x)\right  \},~ \epsilon\gg m_{\pi}.
  \label{eq:eq118}
  \ea
 Note that the expression (\ref{eq:eq112}) is obtained within the assumption of a point (structureless) pion. The effect of the internal structure can be taken into account by introducing a form factor:
 \be
 \eta \to \frac{1}{|F_{\eta}(s)|}\eta.
 \label{eq:eq119}
 \ee
The loop diagrams contribution were calculated supposing a structureless pion.

The charge-even contribution to the cross section, taking into account soft photon emission,
has the expression:
\ba
& -\displaystyle\frac{4\pi\alpha }{16\pi^3}& \int_{\omega<\Delta\epsilon}
 \displaystyle\frac{d^3k}{\omega}\left \{
 \frac{m^2}{(q_{\pm}k)^2};~ \frac{m^2_e}{(p_{\pm}k)^2};~
 \frac{p_+p_-}{p_+k\cdot p_-k};~
 \frac{q_+q_-}{q_+k\cdot q_-k} \right \}= \nn\\
&&
- \displaystyle\frac{\alpha }{\pi}\left \{
\ln  \left (\displaystyle\frac{2\Delta\epsilon}{\lambda}\right ) -
\displaystyle\frac{1}{2\beta}\ln  \frac{1+\beta }{ 1-\beta };
\right . \ln  \left (\displaystyle\frac{2\Delta\epsilon}{\lambda}\right ) -\displaystyle\frac{1}{2}\rho_e;~
\nn\\
&&
\rho_e\ln  \left (\displaystyle\frac{2\Delta\epsilon}{\lambda}\right ) -\displaystyle\frac{1}{4}\rho_e^2 -\frac{\pi^2}{6};
~\displaystyle\frac{1+\beta^2}{ 2\beta }
\left [\ln  \left (\displaystyle\frac{2\Delta\epsilon}{\lambda}\right )
\ln  \frac{1+\beta }{ 1-\beta }-\right .
\nn\\
&&
\displaystyle\frac{1}{4}\ln ^2 \left ( \frac{1+\beta }{ 1-\beta }\right )
+\ln  \frac{1+\beta }{ 1-\beta }\ln  \frac{1+\beta }{ 2\beta }-
\left.  \left.   \frac{\pi^2}{6}+
Li_2\left (\displaystyle\frac{ 1-\beta }{1+\beta }\right )\right ]
\right \}.
\label{eq:eq120}
\ea
From Eqs. (\ref{eq:eq116}) and (\ref{eq:eq120}) one obtains the final result for the charge--even
one--loop contribution to the cross section  (\ref{eq:eq113}). The correction for the emission
of a soft photon takes the following form:
\ba
\displaystyle\frac{d\sigma^{even}}{d\Omega}& =&
\displaystyle\frac{d\sigma_0}{d\Omega}\cdot \frac{2\alpha}{\pi} \left \{ \left (
\rho_e -2 +\displaystyle\frac{1+\beta^2}{ 2\beta }\ln  \frac{1+\beta }{ 1-\beta }\right )
\ln  \left (\displaystyle\frac{\Delta\epsilon}{\lambda}\right )+ \displaystyle\frac{13}{12}\rho_e+
\right . \nn\\
&&
\frac{\pi^2}{6}-\displaystyle\frac{23}{9}+
\displaystyle\frac{1+\beta^2}{ 2\beta }\left [ \rho - \frac{\pi^2}{12}
+Li_2\left (\displaystyle\frac{ 1-\beta }{1+\beta }\right )+
-Li_2\left (-\displaystyle\frac{ 1-\beta }{1+\beta }\right )+
\right  . \nn\\
&&
\displaystyle\frac{3}{2}\ln^2 \left (\displaystyle\frac{1+\beta}{2}\right )-
3\ln \left ( \displaystyle\frac{1+\beta}{2}\right )\ln \beta
-\displaystyle\frac{1}{2}\ln^2\beta +\rho\ln \displaystyle\frac{1+\beta}{ 2\beta^2 }+
\nn\\
&&
\left .3\ln \displaystyle\frac{1+\beta}{2}\right ] +\left .\displaystyle\frac{1-\beta}{ 2\beta }
 \left [\rho +(1+\beta)\ln \displaystyle\frac{1+\beta}{2}\right  ]
 +\Delta_{\mu}+ \Delta_{\pi}
 \right \}.
\label{eq:eq121}
\ea
In the ultrarelativistic limit we obtain:
\be
\displaystyle\frac{d\sigma^{even}}{d\Omega} =
\displaystyle\frac{d\sigma_0}{d\Omega}\cdot \frac{2\alpha}{\pi} \left \{
(\rho+\rho_e-2)\ln\left (\displaystyle\frac{2\Delta\epsilon}{\epsilon}\right )+
\displaystyle\frac{13}{12}\rho_e+\rho+\frac{\pi^2}{12}-\displaystyle\frac{23}{9}
+\Delta_{\mu}+ \Delta_{\pi}
 \right \}.
\label{eq:eq122}
\ee
Let us consider now the case when the additional photon in (\ref{eq:eq114}) is hard.
The corresponding matrix element takes the form :
\ba
{\cal M}^{e^+e^-\to \pi^+\pi^-\gamma}
&=&i(4\pi\alpha)^{3/2}(m_e +m_\pi),\nn\\
m_e&=&\displaystyle\frac{1}{s_1}F_{\pi}(s_1)(q_- - q_+)^{\mu}
\bar v \left [\gamma_{\mu}
\displaystyle\frac{\hat p_- - \hat k}{-2p_-k}\hat e
+\hat e\displaystyle\frac{(-p_+ +k)}{-2p_+k}\gamma_{\mu}
\right ] u ,\nn\\
m_{\pi}&=&\displaystyle\frac{1}{s}F_{\pi}(s)\bar v\gamma_{\mu}u
\left [\displaystyle\frac{(2q_- +k)^{\nu}(k+q_- -q_+)^{\mu}}{2q_-k}+ \right .
\nn\\&&
\left . \displaystyle\frac{(q_--q_+ -k)^{\mu}(-2q_+-k )^{\nu}}{2q_+k}-2q^{\mu\nu} \right ]
e_{\nu}.
\label{eq:eq123}
\ea
It is easy to be convinced that Eqs. (\ref{eq:eq123}) satisfy the relation of gauge invariance:
each term $m_e$ and $m_\pi$, when taken separately, vanishes  under replacement $e \to k$.
The expression (\ref{eq:eq123}) is also in agreement with (\ref{eq:eq116}), when $k\to 0$.
Omitting straightforward calculations, the expression for the cross section of the process
(\ref{eq:eq114}) is:
\be
\sigma^{e^+e^-\to \pi^+\pi^-\gamma}=
\displaystyle\frac{\alpha^3}{2\pi^2 s^2} R \displaystyle\frac{d^3q_+d^3q_-d^3k}
{\epsilon_+\epsilon_-\omega} \delta^{(4)}(p_+ +2p_-- q_+-q_- -k),
\label{eq:eq124}
\ee
where $R=R_0+\Delta R$, with
\ba
R_0&=& \left \{\displaystyle\frac{s}{s_1^2} |F_{\pi}(s_1)|^2
 \left  [ \displaystyle\frac{p_+ p_-}{\chi_+ \chi_-}
 -\displaystyle\frac{2}{\chi_-}-\displaystyle\frac{m_e^2}{\chi^2_-}
 +\displaystyle\frac{\chi_+}{p_+ p_-}
 \left ( \displaystyle\frac{1}{\chi_-}+ \displaystyle\frac{m_e^2}{\chi^2_-}\right ) +
 (q_+ \leftrightarrow q_-) \right ]
\right .+\nn\\
&&\left  .
\displaystyle\frac{2}{s_1}Re[F_{\pi}(s)F_{\pi}^*(s_1)]
\left ( \displaystyle\frac{p_+}{\chi_+}- \displaystyle\frac{p_-}{\chi_-}\right )
\left ( \displaystyle\frac{q_+}{\chi_+^{'}}- \displaystyle\frac{q_-}{\chi_-^{'}}\right )
\right \} (p_+ Qp_-Q),
\label{eq:eq125a}
\ea
\ba
\Delta R&=&\displaystyle\frac{s}{s_1^2}|F_{\pi}(s_1)|^2
 \left  [ \displaystyle\frac{1}{\chi_-}(p_+ Qk_-Q)+
 \displaystyle\frac{1}{\chi_+}(p_- Qk_+Q)+
 \displaystyle\frac{2Qk}{\chi_+\chi_-}(p_+ Qp_-k)\right ]+
 \nn\\
&&
 \displaystyle\frac{1}{s}|F_{\pi}(s)|^2  \left \{
 \displaystyle\frac{ q_+q_-}{\chi_+^{'}\chi_-^{'}}[ \chi_+\chi_- -2(p_+ Q p_-q)]+
 \displaystyle\frac{m^2}{4\chi_+^{'2}} (p_+(k+2Q)p_-(k+2Q))-\right .
 \nn\\
&&
 \left  .\displaystyle\frac{1}{\chi_+^{'}} [\chi_+\cdot p_-q_+ +\chi_ -\cdot p_+q_+
+2(p_-Qp_+q_+)+ p_+p_-\cdot Q^2] +(q_+ \leftrightarrow q_-) \right \}+
 \nn\\
&&
\displaystyle\frac{1}{s_1}Re[F_{\pi}(s)F_{\pi}^*(s_1)]
\left \{  (p_+ Qp_-k)
\left ( \displaystyle\frac{p_+}{\chi_+}-
\displaystyle\frac{p_-}{\chi_-}\right ) \displaystyle\frac{q_+}{\chi_+^{'}}\right .
-2(p_+ -p_-)Q-
\nn\\
&&
 \displaystyle\frac{1}{\chi_+^{'}}(k(p_+-p_-) Qq_+)
+\displaystyle\frac{2Qk\cdot Qp_+\cdot p_-q_+}{\chi_-\chi_+^{'}}-
 \displaystyle\frac{2Qk\cdot Qp_-\cdot p_+q_+}{\chi_+\chi_+^{'}}+
\nn\\
&&
 \left .Q^2
 \left [ -\displaystyle\frac{\chi_+\cdot q_+p_-}{\chi_-\chi_+^{'}}+
 \displaystyle\frac{\chi_-\cdot q_+p_+}{\chi_+\chi_+^{'}}+
 \displaystyle\frac{q_-(p_+-p_-)}{\chi_+^{'}}\right ]
-(q_+ \leftrightarrow q_-) \right \},\label{eq:eq125b}
\ea
and
$$Q=\displaystyle\frac{1}{2} (q_+ -q_-),~(abcd)=ab\cdot cd+ad\cdot bc-ac\cdot bd,
~k_\pm=k-p_\pm\displaystyle\frac{kp_\mp}{p_+p_-}.$$
The quantity $\Delta R$ does not contain collinear divergence, i.e. it is finite when $\chi_{\pm}\to 0$. In the ultrarelativistic limit we obtain
\ba
R&=& |F_{\pi}(s)|^2
\left  [ \displaystyle\frac{tu+t_1u_1}{ss_1}\displaystyle\frac{4s_1}{\chi_+^{'}\chi_-^{'}}-
\displaystyle\frac{8m^2}{s^2}\left (
\displaystyle\frac{tu_1}{\chi_+^{'2}}+
\displaystyle\frac{t_1u}{\chi_-^{'2}} \right)\right ] +\label{eq:eq126}\\
&&
|F_{\pi}(s_1)|^2\left  [ \displaystyle\frac{tu+t_1u_1}{ss_1}
\displaystyle\frac{4s}{\chi_+\chi_-}-
\displaystyle\frac{4m_e^2}{ss_1}
(tu_1+t_1u )\left (\displaystyle\frac{1}{\chi_+^2}+
\displaystyle\frac{1}{\chi_-^2}+ \right)\right ] +\nn\\
&&
\displaystyle\frac{[tu+t_1u_1]}{ss_1}
\displaystyle\frac
{[ss_1(t+t_1-u-u_1)+(s+s_1)(tt_1-uu_1)]}
{\chi_+\chi_+^{'}\chi_-\chi_-^{'}}
Re[F_{\pi}(s)F_{\pi}^*(s_1)],
\nn
\ea
where
\ba
s&=& (p_+ + p_-)^2,~s_1= (q_+ + q_-)^2,~
t= (p_- - q_-)^2,~t_1= (p_+ - q_+)^2, \nn\\
u&=& (p_- - q_+)^2,~u_1= (p_+ - q_-)^2,~
\chi_{\pm}=k p _{\pm},~\chi_{\pm}^{'}=k q_{\pm}.
\label{eq:eq127}
\ea

\section{ The process $e^++e^-\to\mu^++\mu^- (\gamma)$}

\begin{figure}
\begin{center}
\includegraphics[width=15cm]{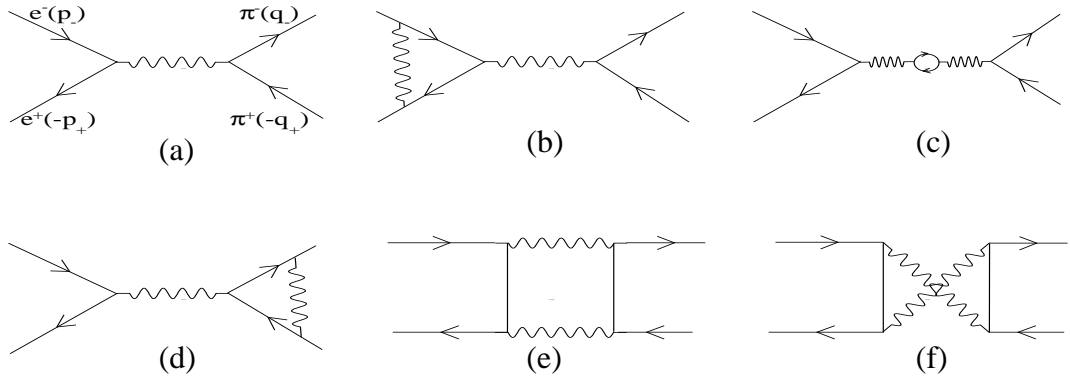}
\caption{\label{Fig:fig3}Feynman Diagrams which contribute to the charge--even part of the cross section for the process $e^++e^-\to  \pi^++\pi^-(\gamma)$.}
\end{center}
\end{figure}

We limit our considerations to the energy region where one can neglect large corrections to the calculation of
the cross section, due to the intermediate state with exchange of a particle of mass $M_l$ such that:
\be
\displaystyle\frac{s}{M_l^2}\le 10^{-3}.
\label{eq:eq21}
\ee
Therefore we do not consider the $Z$-boson contribution.
Part of the following result has already been given earlier in the literature, but it is rederived it here for completeness.
The cross section of the process $e^+(p_+) +e^-(p_-)\to\mu^+(q_+)+\mu^-(q_-) $ in the Born approximation can be written as:
\ba
\displaystyle\frac{d\sigma_0}{d\Omega}&=&
\displaystyle\frac{\alpha^2 \beta}{4s}(2-\beta^2\sin^2\theta),~\beta=\left (1-\displaystyle\frac{m_\mu^2}{\epsilon^2}\right )^{1/2},~s=4\epsilon^2,~\theta=\widehat{\vec q_-,\vec p_-}.
\label{eq:eq22}
\ea
Let us look the the charge--even part of the cross section, determined by the Feynman diagrams Figs. \ref{Fig:fig3}a-e.

In the case of $\beta\sim 1$, the contribution of the muon form factor $F_2$ must be taken into account, too:
$$
\Gamma^{\mu}(q)= \gamma^{\mu}F_1(q^2)-\displaystyle\frac{1}{4m}[\gamma_{\mu},~(\hat q_++\hat q_-)] F_2(q^2),~ F_1(q^2)=1+F_1^{(2)}(q^2),~F_1^{(2)}(0)=0.$$
Let us start from the interference with the Born amplitude
\ba
& \displaystyle\frac{1}{4} Tr \hat  p_+ \gamma^{\mu} \hat  p_-  \gamma^{\nu}\bigotimes
&\displaystyle\frac{1}{4} Tr (\hat q_-+m)\left (  \gamma_{\mu} F_1 -
  \displaystyle\frac{[ \gamma_{\mu},(\hat q_++\hat q_-)]}{4m}F_2 \right )
  (\hat q_+ -m)
\nn\\
&&
\left (  \gamma_{\mu} F_1 +
  \displaystyle\frac{[ \gamma_{\nu}(\hat q_++\hat q_-)]}{4m}F_2 \right )=
\nn\\
&&
 =\displaystyle\frac{s^2}{4}(2-\beta^2\sin^2\theta)
 \left [F_1^2+  \displaystyle\frac{4}{2-\beta^2\sin^2\theta }F_2 F_1\right ]
 \approx\nn\\
&&
 \approx
  \displaystyle\frac{s^2}{4}(2-\beta^2\sin^2\theta)
\left [1+  2F_1^{(2)} + \displaystyle\frac{4}{2-\beta^2\sin^2\theta }F_2\right].
\nn
\ea
Introducing the the expression of the form factors:
 \ba
 F_1^{(2)} &=& \frac{\alpha}{\pi}
 \left \{
 \left (\ln \displaystyle\frac{m}{\lambda} -1\right )
 \left (1- \displaystyle\frac{1+\beta^2}{ 2\beta }
 \ln \displaystyle\frac{1+\beta}{1-\beta} \right )+
 \right  .\nn\\
&&
   \displaystyle\frac{1+\beta^2}{2\beta}
 \left [ \displaystyle\frac{\pi^2}{3}- \displaystyle\frac{1}{4}\ln^2
 \left (  \displaystyle\frac{1+\beta}{1-\beta}\right )+
 \right  .
 \nn\\
&&
\left .\left .
\ln\left (  \displaystyle\frac{1+\beta}{1-\beta}\right )
\ln\left (  \displaystyle\frac{1+\beta}{2\beta}\right )
+Li_2\left (  \displaystyle\frac{1-\beta}{1+\beta} \right )\right ]
-\displaystyle\frac{1}{4\beta}\ln\left (  \displaystyle\frac{1+\beta}{1-\beta}\right )\right \} ,
 \nn\\
F_2&=&\frac{\alpha}{\pi}\cdot \displaystyle\frac{1-\beta^2}{4\beta } \ln \left (\displaystyle\frac{1-\beta}{1+\beta}\right ),
 \nn
\ea
and deriving the contribution of soft photon emission (with the same precision with
respect to the change $\beta_\pi\to \beta_\mu\equiv \beta$) similarly to the case of pion production:
\ba
\left( \displaystyle\frac{d\sigma}{d\Omega}\right ) ^{soft}_{even}&=&
\displaystyle\frac{2\alpha }{\pi}\displaystyle\frac{d\sigma_0}{d\Omega}
\left \{ (\rho_e-1)\ln \left(\displaystyle\frac{2\Delta\epsilon}{\lambda}\right )
-\displaystyle\frac{1}{4}\rho_e^2 - \displaystyle\frac{\pi^2}{6}+\displaystyle\frac{1}{2}\rho_e+
\right .
\nn\\
&&
\displaystyle\frac{1}{ 2\beta }\ln  \frac{1+\beta }{ 1-\beta } +
\ln \left(\displaystyle\frac{2\Delta\epsilon}{\lambda}\right )
 \left(\displaystyle\frac{1+\beta^2}{ 2\beta }\ln  \displaystyle\frac{1+\beta }{ 1-\beta }-1\right )+
\nn\\
&&
\displaystyle\frac{1+\beta^2}{ 2\beta }
\left[-\displaystyle\frac{1}{4}\ln ^2 \left(\displaystyle\frac{1+\beta }{ 1-\beta }\right )
+\ln  \displaystyle\frac{1+\beta }{ 1-\beta }
\ln  \displaystyle\frac{1+\beta }{ 2\beta }
- \frac{\pi^2}{6 } +\right .
\nn\\
&&
\left .\left .
Li_2 \left(\displaystyle\frac{1-\beta }{ 2\beta }\right )\right ]\right \},  \label{eq:eq23}
\ea
we obtain the following expression for the charge--even contribution to the cross section:
 \ba
 \left( \displaystyle\frac{d\sigma}{d\Omega}\right )_{even}&=&
\displaystyle\frac{2\alpha }{\pi}\displaystyle\frac{d\sigma_0}{d\Omega}
\left \{ \left (\rho_e-2+ \displaystyle\frac{1+\beta^2}{ 2\beta }\ln
\displaystyle\frac{1+\beta }{ 1-\beta }\right )\ln \left(\displaystyle\frac{\Delta\epsilon}{\epsilon}\right ) -\right .
\nn\\
&&
(1-\beta^2)\ln \left ( \displaystyle\frac{1+\beta }{ 1-\beta }\right )
[2\beta(2-\beta^2\sin^2\theta)]^{-1}+
\nn\\
&&
\displaystyle\frac{13}{12}\rho_e +\displaystyle\frac{\pi^2}{6}
 -\displaystyle\frac{31}{9}+ \displaystyle\frac{1}{3}\beta^2
+\rho \left (-\displaystyle\frac{1}{2}+ \displaystyle\frac{3}{4\beta}+\beta - \displaystyle\frac{1}{6}\beta^3 \right )+
\nn\\
&&
\left (\displaystyle\frac{1}{2\beta}+2\beta - \displaystyle\frac{1}{3}\beta^3 \right ) \ln  \displaystyle\frac{1+\beta }{ 2 }
+ \displaystyle\frac{1+\beta^2 }{ 2\beta }
\left  [ \displaystyle\frac{\pi^2}{6}+ 2\rho \ln  \displaystyle\frac{1+\beta }{ 2\beta }
+ \right .
\nn\\
&&
\left .\left .
 2 \ln  \displaystyle\frac{1+\beta }{ 2} \ln  \displaystyle\frac{1+\beta }{ 2\beta^2 }
+2Li_2\left(\displaystyle\frac{1-\beta }{ 1+\beta }\right )\right ]+\Delta_H \right \}.
\label{eq:eq24}
\ea
 Let us note that Ref. \cite{Be73} contains a mistake: a factor two is missing in the second term in parenthesis of Eq. (\ref{eq:eq24}).

In the ultrarelativistic limit we obtain:
 \ba
 \left( \displaystyle\frac{d\sigma}{d\Omega}\right )_{even}&=&
\displaystyle\frac{\alpha^3 }{2\pi s} (1+c^2)
\left [ ( \rho_e + \rho -2)  \left ( \ln \displaystyle\frac{\Delta\epsilon}{\epsilon}
 +\displaystyle\frac{13}{12} \right )+\displaystyle\frac{\pi^2}{3}-
\right . \nn\\
&&
 \left .
 \displaystyle\frac{17}{18} +\Delta_H\right ].
 \label{eq:eq25}
\ea
Let us consider now the charge--odd part of the reaction $e^+ +e^-\to\mu^++\mu^-$.
The one--loop contribution is described by the diagrams \ref{Fig:fig3}d--e and has the form \cite{Be73,Ku77}
\ba
\left( \displaystyle\frac{d\sigma}{d\Omega}\right ) ^{virt}_{odd}&=&
\displaystyle\frac{\alpha^3\beta }{2\pi s}\left \{
(2-\beta^2\sin^2\theta)\ln  \displaystyle\frac{1+\beta c}{ 1-\beta c }
\ln \left( \displaystyle\frac{2\epsilon}{\lambda}\right )-
\right .
\nn \\
&&
  \displaystyle\frac{1-\beta^2(1+\sin^2\theta)}{ 1+\beta^2-2\beta c }
(\rho +\ell_-)+
\beta c \left [ - \displaystyle\frac{1}{2\beta^2} \rho -\displaystyle\frac{1}{2} s F_{\Delta}+
\right .
\nn \\
&&
\displaystyle\frac{1}{2} s F_Q \left (-1 - \displaystyle\frac{\beta^2}{2} + \displaystyle\frac{1}{2\beta^2}\right ) +
\displaystyle\frac{1}{4} ( \rho_e^2 + \rho^2)
+\displaystyle\frac{\pi^2}{6}-\displaystyle\frac{1}{2} \ell_-^2 +
\nn \\
&&
\left .
L_-( \rho+ \ell_-)
-Li_2  \left( \displaystyle\frac{1-\beta^2}{2(1-\beta c) } \right )\right ] -
(1-\beta^2)\left [\displaystyle\frac{1}{4} \ell_-^2-\right .\nn\\
&&\left .
\left .
\displaystyle\frac{1}{2} L_- ( \rho+ \ell_-)+\displaystyle\frac{1}{2 } Li_2 \left(\displaystyle\frac{1-\beta^2}{2(1-\beta c)} \right )\right ]
 - (c\leftrightarrow -c)\right \}.
\label{eq:eq26}
\ea
All quantities, entering in (\ref{eq:eq26}), are defined above (see Eqs. (\ref{eq:eq15},\ref{eq:eq18})). The corresponding contribution of soft photon emission  (within the precision in permuting $\beta_{\pi}\to \beta_{\pi}=\beta $ with the corresponding one for the process $e^++ e^-\to\pi^++\pi ^-$) has the form:
\ba
\left( \displaystyle\frac{d\sigma}{d\Omega}\right ) ^{soft}_{odd}&=&
\displaystyle\frac{\alpha^3\beta }{2\pi s}( 2-\beta^2 \sin^2\theta)
\left \{
-\ln  \displaystyle\frac{1+\beta c}{ 1-\beta c }
\ln \left( \displaystyle\frac{2\Delta\epsilon}{\lambda}\right )+ 
\displaystyle\frac{1}{2}\ell_-^2 -
\right .\nn\\
&&
L_-( \rho+ \ell_-)+Li_2 \left(\displaystyle\frac{1-\beta^2}{2(1-\beta c)} \right )+
Li_2 \left(\displaystyle\frac{\beta^2(1-c^2)}{ 1+\beta^2-2\beta c }\right )-
\nn\\
&&
\left .
\int _0^{1-\beta^2} \displaystyle\frac{dx}{x}
\left (1-\displaystyle\frac{x}{\chi_t }\right )^{-1/2} f(x)-(c\leftrightarrow -c)\right \}
\label{eq:eq27}
\ea
Finally
 we derive the expression for the charge--odd part of the cross section $(d\sigma/d\Omega)^{odd}$
and for the related coefficient of the charge asymmetry,
\be
\eta=\left( \displaystyle\frac{d\sigma}{d\Omega}\right )_{odd}/\left(
\displaystyle\frac{d\sigma_0}{d\Omega}\right ),
\label{eq:eq28}
\ee
where
\ba
\left( \displaystyle\frac{d\sigma}{d\Omega}\right )_{odd}&=&
\displaystyle\frac{\alpha^3 \beta}{2\pi s}\left \{
(2-\beta^2\sin^2\theta)
\left [\ln  \displaystyle\frac{1+\beta c}{ 1-\beta c }
\ln \left( \displaystyle\frac{\epsilon}{\Delta\epsilon}\right )
\displaystyle+\frac{1}{2}\ell_-^2 -
\right .\right .
\nn\\
&&
 L_-( \rho+ \ell_-)+Li_2 \left(\displaystyle\frac{1-\beta^2}{2(1-\beta c)} \right )+
Li_2 \left(\displaystyle\frac{\beta^2(1-c^2)}{ 1+\beta^2-2\beta c }\right )
-\nn\\
&&
\left .
\int _0^
{1-\beta^2} \displaystyle\frac{dx}{x}f(x)\left (1-\displaystyle\frac{x}{\chi_t }\right )^{-1/2}
\right ]
-\displaystyle\frac{ 1-\beta^2 (1+\sin^2\theta)}{ 1+\beta^2-2\beta c }( \rho+ \ell_-)-
\nn\\
&&
\displaystyle\frac{1}{4}(1-\beta^2)
\left [\ell_-^2 -2L_-( \rho+ \ell_-)+
2Li_2 \left(\displaystyle\frac{1-\beta^2}{2(1-\beta c)} \right )\right ] +\nn\\
&&
\beta c\left[
-\frac{1}{2\beta^2}\rho+\frac{\pi^2}{12}+\frac{1}{4}\rho^2+\displaystyle\frac{1
}{2\beta}\left( -1 - \displaystyle\frac{ \beta^2}{2} +
\displaystyle\frac{1}{2\beta^2}\right )
\right .
\nn\\
&&
\left(\displaystyle\frac{\pi^2}{6}
+ \rho\ln  \displaystyle\frac{1+\beta }{ 1-\beta  }
-\displaystyle\frac{1}{2}\ln ^2 
\left(\displaystyle\frac{1+\beta }{ 1-\beta  }\right )-
4Li_2\left(\displaystyle\frac{1-\beta }{ 2 }\right ) -
\right .
\nn\\
&&
\left . 2Li_2\left(-\displaystyle\frac{1-\beta }{1+\beta}\right ) -2\ln \displaystyle\frac{1-\beta }{ 2 }\ln \displaystyle\frac{1+\beta }{ 2 } \right ) -\displaystyle\frac{1}{ 2 }\ell_-+
\nn\\
&&
\left .\left .
L_-( \rho+ \ell_-)-Li_2 \left(\displaystyle\frac{1-\beta^2}{2(1-\beta c)} \right )\right ]
 - (c\leftrightarrow -c)\right \}.
\label{eq:eq29}
\ea
In the ultrarelativistic limit this relation coincides with the result firstly derived by I. B.  Kriplovitch\cite{Kr73} :
\ba
 \left( \displaystyle\frac{d\sigma}{d\Omega}\right )^{as}_{odd}&=&
\displaystyle\frac{\alpha^3 }{\pi s}
\left \{ 2(1+\cos^2\theta)\left [\ln\cot \displaystyle\frac{\theta}{2}\ln \left( \displaystyle\frac{\epsilon}{\Delta\epsilon}\right )+\right .
\displaystyle\frac{1}{2}\ln ^2\sin\displaystyle\frac{\theta}{2}-\right .
\label{eq:eq210}\\
&&
\left .
\displaystyle\frac{1}{2}\ln ^2\cos\displaystyle\frac{\theta}{2}+
\displaystyle\frac{1}{4}
\int_{\cos^2\theta/2}^{\sin^2\theta/2}\displaystyle\frac{dx}{x}\ln (1-x)
\right ]+
\cos^2\displaystyle\frac{\theta}{2}\ln \sin\displaystyle\frac{\theta}{2}-
\nn\\
&&
\left .
\sin^2\displaystyle\frac{\theta}{2}\ln \cos\displaystyle\frac{\theta}{2}-
 \left(\ln ^2\cos\displaystyle\frac{\theta}{2}+\ln ^2\sin\displaystyle\frac{\theta}{2}\right )
 \cos\theta \right \},~ \theta=\widehat{\vec q_-,\vec p_-}.
\nn
\ea
\subsection{The process  $e^+(p_+) +e^-(p_-)\to\mu^+(q_+)+\mu^-(q_-) +\gamma(k)$}
The cross section of this process has the following form \cite{Be73,Ku77}:
\ba
 d\sigma &=&
 \displaystyle\frac{\alpha^3 }{2\pi^2 s^2} R
 \displaystyle\frac{d^3q_+d^3q_-d^3k}
{\epsilon_+\epsilon_-\omega} \delta^{(4)}(p_+ +p_-- q_+-q_- -k),
\label{eq:eq211}
\nn\\
R&=&- \displaystyle\frac{1}{2} s (uu_1+t t_1)a^2+
a\left (r+ \displaystyle\frac{s}{s_1} q\right )+
\displaystyle\frac{2}{s_1}(t+t_1-u-u_1)-
\nn\\
&&
\displaystyle\frac{1}{s\chi_-^{'}}
(t_1\chi_-+u\chi_+) - \displaystyle\frac{1}{s\chi_+^{'}}(u_1\chi_-+t_1\chi_+)-\nn\\
&&
\displaystyle\frac{s}{s_1^2\chi_-}(u_1\chi_+^{'}+t_1\chi_-^{'}) -
\displaystyle\frac{s}{s_1^2\chi_+}(u\chi_-^{'}+t_1\chi_+^{'}),
\nn
\ea
where the determination of the invariants coincides with (\ref{eq:eq127}). Moreover the four vectors $a$, $r$, $q$ can be expressed as:
\ba
a&=&\displaystyle\frac{1}{s} (n^{'}_--n^{'}_+)+ \displaystyle\frac{1}{s} (n_+-n_-),
n_{\pm}=\displaystyle\frac{p_{\pm}}{\chi_{\pm}},n^{'}_{\pm}=\displaystyle\frac{p_{\pm}}{\chi^{'}_{\pm}},\nn\\
q&=&q_-(u-t_1)+q_+(t-u_1)+ n_-(\chi_+^{'}u_1+\chi_-^{'}t_1)- n_+(\chi_-^{'}u_1+\chi_+^{'}t),
\nn\\
r&=&p_+(t-u)+p_-(u_1-t_1)+ n_-^{'}( \chi_-t_1+\chi_+u)-n_+^{'}(\chi_- u_1+\chi_+ t).
\label{eq:eq212}
\ea
In the ultrarelativistic limit $1-\beta_\mu\ll 1$,
the expression for the cross section takes the form:
\ba
d\sigma&=& \displaystyle\frac{\alpha^3 }{2\pi^2 s} X d\Omega_{\mu} d q_+^0 d q_-^0d \phi_{\gamma}
= \displaystyle\frac{\alpha^3 }{2\pi^2 s} \displaystyle\frac{|q_+|\omega}{2\epsilon-\omega( 1-\cos\theta_{\gamma})} X d\Omega_{\mu} d\Omega_{\gamma}d\omega,\nn\\
X&=& -\displaystyle\frac{m_e^2}{2s_1^2}
\left (\displaystyle\frac{t_1^2+u_1^2}{\chi_-^2}+\displaystyle\frac{t^2+u^2}{\chi_+^2}\right )-
\displaystyle\frac{m_{\mu}^2}{2s^2}
\left (\displaystyle\frac{t_1^2+u^2}{\chi^{'2}_-}+
\displaystyle\frac{t^2+u_1^2}{\chi^{'^2}_+}\right )+
 \nn\\
&&
(t ^2+t_1^2+u^2+u_1^2)\left [\displaystyle\frac{1}{4s_1\chi_+\chi_-}+ 
\displaystyle\frac{1}{4s\chi_+^{'}\chi_-^{'}}+ \right .
\nn\\
&&
\left .\displaystyle\frac{1}{4ss_1}\left (\displaystyle\frac{ u_1}{\chi_+\chi^{'}_-}
+ \displaystyle\frac{ u}{\chi_-\chi^{'}_+}
- \displaystyle\frac{ t_1}{\chi_+\chi^{'}_+}
- \displaystyle\frac{ t}{\chi_-\chi^{'}_-}\right )\right ].
\label{eq:eq213}
\ea
\section{The process $e^+ +e^- \to e^+ +e^-(+\gamma)$}

\subsection{Bhabha scattering: $e^+ +e^- \to e^+ +e^-$}

The expression for the cross section of Bhabha scattering:
$e^+(p_+) +e^-(p_-) \to e^+(q_+) +e^-(q_-)$
taking into account the correction for the emission of soft photon, in the one--loop approximation takes the form
\be
\displaystyle\frac{d\sigma}{d\Omega}=
\displaystyle\frac{d\sigma_0}{d\Omega}( 1+\delta),~
\displaystyle\frac{d\sigma_0}{d\Omega}=
\displaystyle\frac{\alpha^2}{4s}\left (\displaystyle\frac{3+c^2}{1-c}\right )^2,~c=\cos\theta=\cos\widehat{\vec q_-,\vec p_-}.
\label{eq:eq31}
\ee
\ba
\delta &=&  \displaystyle\frac{2\alpha}{\pi}
\left [ 2\left (1-\rho+2\ln \cot\frac{\theta}{2}\right )
\ln\frac{\epsilon}{\Delta\epsilon}+
\int ^{\sin^2(\theta/2)}_{\cos^2(\theta/2)}\frac{dx}{x}\ln(1-x)-
\right .
 \nn\\
 &&
 \left .
 \displaystyle\frac{23}{9}+\displaystyle\frac{11}{6}\rho \right  ]+
  \displaystyle\frac{\alpha }{\pi}(3+c^2)^{-2} 
 \left \{ \displaystyle\frac{\pi^2}{3}(2c^4-3c^3-15c)
 +
 \right . \nn\\
 &&
 2(2c^4-3c^3+9c^2+3c+21) \ln^2\sin\frac{\theta}{2}-
 \nn\\
 &&
 4c(c^4+c^2-2c)\ln^2\cos\frac{\theta}{2}
 -4(c^3+4c^2+5c+6)\ln^2\left (\tan\frac{\theta}{2}\right )+
 \nn\\
 &&
 \displaystyle\frac{2}{3}(11c^3+33c^2+21c+111) \ln\sin\frac{\theta}{2}+
 2(c^3-3c^2+7c-5)\ln\cos\frac{\theta}{2}+\nn\\
 &&
 2(c^3+3c^2+3c+9) \delta_t -2(c^3+3c)(1-c) \delta_s \Big \},
 \nn
 \ea
with
$$
 \delta_s=(\Delta_{\mu}+\Delta_ H),~ \delta_t=(\Delta_{\mu}+\Delta_ H)_{s\to -\frac{s}{2}(1-c)} .
$$

\subsection{The process $e^+(p_+) +e^-(p_-) \to e^+(q_+) +e^-(q_-)+\gamma(k)$}
The cross section of the process $e^+(p_+) +e^-(p_-) \to e^+(q_+) +e^-(q_-)+\gamma(k)$ was 
firstly obtained in Ref. \cite{Be82} in a compact form:
$$
d\sigma^{e^+e^-\to e^+e^-\gamma}=
\displaystyle\frac{\alpha^3}{8\pi s} \Gamma_{e^+e^-}
\displaystyle\frac{d^3q_+d^3q_-d^3k}
{q_+^0q_-^0\omega} \delta^{(4)}(p_+ +p_-- q_+-q_- -k),
$$
\ba
\Gamma_{e^+e^-}&=&W
\displaystyle\frac{[ss_1(s^2+s_1^2)+tt_1(t^2+t_1^2)+uu_1(u^2+u_1^2)]}{ss_1tt_1}
-
\nn\\
&&
\displaystyle\frac{4m^2}{\chi^{'2}_+}
\left (\displaystyle\frac{s }{t }+\displaystyle\frac{t }{s}+1\right )^2-
\displaystyle\frac{4m^2}{\chi^{'2}_-}
\left (\displaystyle\frac{s_1 }{t_1 }+\displaystyle\frac{t_1 }{s_1}+1\right )^2
-
\nn\\
&&
\displaystyle\frac{4m^2}{\chi^{2}_+}
\left (\displaystyle\frac{s_1 }{t }+\displaystyle\frac{t }{s_1}+1\right )^2-\displaystyle\frac{4m^2}{\chi^{2}_-}
\left (\displaystyle\frac{s }{t_1 }+\displaystyle\frac{t_1 }{s}+1\right )^2,\label{eq:eq32}
\\
W&=&\displaystyle\frac{1}{4\chi_+\chi_+^{'}\chi_-\chi_-^{'}}
[u(st+s_1 t_1 )+u_1(s_1t+s t_1 )+ 2ss_1(t+t_1)+\nn\\
&&
2tt_1(s+s_1)]= - \left (\displaystyle\frac{p_-}{\chi_-}- \displaystyle\frac{p_+}{\chi_+}
\left .-\displaystyle\frac{q_+}{\chi_+^{'}}+\displaystyle\frac{q_-}{\chi_-^{'}}\right )^2
\right |_{m_e^2=0},
\nn
\ea
where the invariants were previously defined in (\ref{eq:eq127}).
Below, we will try to generalize the results obtained in (\ref{eq:eq31}) and (\ref{eq:eq32}) taking into account higher order corrections. Conserving in (\ref{eq:eq31}) only the terms of the sum of the order
${\alpha}/\pi\rho\simeq 1$ (which do not vanish in the limit (\ref{eq:eq2})), 
we rewrite Eq. (\ref{eq:eq31}) as:
\be
d\sigma=d\sigma_0\left \{ 1+\beta \left [ 2 \ln\frac{\Delta\epsilon}{\epsilon}+\displaystyle\frac{11}{6}\right ]\right \},~
\beta=\frac{2\alpha}{\pi}(\rho-1)\le 1.
\label{eq:eq33}
\ee
This result can be reproduced in a general form, as an exact function of the $\beta$ parameter, with the help of the formalism of the renormalization group (8).
It is possible to see, indeed, that the functions
\be
{\cal D}(x,\beta_q)= \displaystyle\frac{\beta_q}{2}
\left [ (1-x)^{\beta_q/2-1} \left (1+\frac{3}{8}\beta_q\right) -
\frac{1}{2}(1+x)\right ],
\label{eq:eq34}
\ee
with
$$
\beta_q= \displaystyle\frac{2\alpha}{\pi}
\left (\ln \left |\frac{q^2}{m^2}\right |-1\right ), ~\int_0^1dx{\cal D}(x,\beta_q)=1,
$$
describe the (parton--like) probability to find, in the initial electron (positron), an electron (positron) with an energy fraction $x$ of the initial
energy and four--momentum squared $|q^2|\le s$, and write the cross--section of the process $e^++e^-\to e^{+'}+e^{-'}+...$ in the form of a Drell--Yan process:
 \ba
 &&d\sigma(\epsilon n_-, \epsilon n_+;\epsilon yn'_-,
 \epsilon zn'_+)=
 \int^1 dx_1\int^1 dx_2 {\cal D}(x_1,\beta_q){\cal D}(x_2,\beta_q) d\sigma_0\nn\\
 &&
 ( \epsilon x_1n_-, \epsilon x_2n_+;\epsilon y_1n'_-,\epsilon y_2 n'_+)
 \int^1\displaystyle\frac{dy}{y_1}{\cal D}\left (\displaystyle\frac{y}{y_1},\beta_q \right )
\int^1\displaystyle\frac{dz}{y_2}{\cal D}\left (\displaystyle\frac{z}{y_2},\beta_q\right ) 
 \nn\\
 && 
 (1-\Pi)^{-2},
 \label{eq:eq35}
\ea
where $n_{\pm}$, $n'_{\pm}$ are unit four--momenta of the initial and final $e^+$, $e^-$. The result 
(\ref{eq:eq33}) is obtained by setting $(1-\Pi)^{-2}=[1-(\alpha/(3\pi)\rho]^{-2}$ and choosing 
the low limit of the integration over the energy fraction $y_0=x_0=[1-(\Delta\epsilon/\epsilon)]$, 
where $\Delta\epsilon$ corresponds to the precision of the energy measurement of the electron and 
the positron.

It is necessary to point out, that the description of the process in frame of the partonic 
model (\ref{eq:eq35}) is only true for a quasi--two--particle kinematics: when the final particles 
group themselves into a jet moving along the beam axis (emitted by the initial $e^+$, $e^-$) or 
along the direction  of the final electron and positron, scattered at large angles.
In such case, the cross section appears to be a function of the parameter  $(\alpha/\pi)\rho$, 
(\ref{eq:eq2}): (we set $\ln(\Delta\epsilon/\epsilon)\sim 1$, $\ln(\Delta\theta/\theta)\sim 1$) 
and it can be calculated from Eq. (\ref{eq:eq35}). It is known that the cross section for 
$e^+e^-$ scattering in a particular kinematics (forward or backward scattering) appears as a 
function of the parameter  $(\alpha/\pi)\rho^2$ \cite{AB81} and can not be described by a formula 
of type (\ref{eq:eq35}).

We note that here we neglected the effect of the formation of an additional pair  $e^+e^-$ 
(it represents less than 0.2\% in the energy range 200 MeV$\le \sqrt{s}\le$ 3 GeV, and, moreover it 
can be calculated.  In most of the experiments, the events with pair production are rejected, which 
allows us to write a simple interpolated formula (\ref{eq:eq35}) for ${\cal D}(x,\beta_q)$.
In the leading logarithm approach  (\ref{eq:eq2}), $(\alpha/\pi)\rho\sim 1$, the initial electron 
and positron keep their direction, up to the momentum of the process of hard scattering at large 
angle. This last emission also does not change their direction, defined in the subprocess of hard 
scattering of partons: the electrons with energy fraction $x_1$ and the positrons with energy 
fraction $x_2$ going to electrons  with energy fraction $y_1$ emitted at an angle $\theta_-$ with 
respect to $\vec p_-$ and to positrons with energy fraction $y_2$, scattered at an angle $\theta_+$ 
with respect to $\vec p_-$. Applying conservation laws for the subprocesses, one finds  the following relations:
\be
x_1+x_2=y_1+y_2, ~y_1\sin\theta_-=y_2\sin\theta_+,~ x_1-x_2=y_1\cos\theta_-+y_2\cos\theta_+.
\label{eq:eq36}
\ee
In Eq.  (\ref{eq:eq36}) we use the fact, characteristic for the soft process, that in  Eq.  
(\ref{eq:eq2})  the emission does not change the direction of motion of the partons, which means 
that the three--momentum of the emitted $e^+$, $e^-$ lies in the plane containing the axis of the 
beams, i.e., the angle between the planes $\vec p '_- \vec p_-$ and $\vec p'_+ \vec p_-$ is 180$^0$. 
When the energies and scattering angles of electrons and positrons are measured, Eq. (\ref{eq:eq35})
can be written in the following form (with the help of (\ref{eq:eq36})):
\ba
\displaystyle\frac{d\sigma^{e^+e^-\to e^{+'}e^{-'}+...}}{d\theta_-\theta_+dydz}&=&
\displaystyle\frac{\pi\alpha^2(2-2c_+c_-+s_+s_-)^2}{(1-c)(1-c_-)^2(1-c_+^2)\cdot s}
\int^1\displaystyle\frac{dx}{x_1^3}{\cal D}(x_1,\beta)
\label{eq:eq37}\\
&&
{\cal D}(x_2,\beta)
{\cal D}(\displaystyle\frac{y}{y_1},\beta)
{\cal D}(\displaystyle\frac{z}{y_2},\beta) \cdot K\cdot (1-\Pi)^{-2},
\nn
\ea
with
$$
x_2=x_1\tan \displaystyle\frac{\theta_-}{2}\tan\displaystyle\frac{\theta_+}{2},~
~\theta=\theta_++\theta_-,$$
$$
y_1=x_1 \displaystyle\frac{\sin \displaystyle\frac{\theta_+}{2}}{ \sin \displaystyle\frac{\theta}{2}\cos  \displaystyle\frac{\theta_-}{2}},~
y_2=x_1\displaystyle\frac{\sin \displaystyle\frac{\theta_-}{2}}{ \sin \displaystyle\frac{\theta}{2}\cos  \displaystyle\frac{\theta_+}{2}}.
$$
$\epsilon y=\epsilon _-$,~$\epsilon z=\epsilon _+$ are the energies of the detected electron and positron, $\theta_{\pm}$ the angles of emission with respect to the direction of $\vec p_-$, $c=\cos\theta$,  $c_{\pm}=\cos\theta_{\pm}$, $s_{\pm}=\sin\theta_{\pm}.$

The quantity $K$ in Eq. (\ref{eq:eq37}) is called K-factor, which can be calculated asymptotically for the non leading terms, as it does not contain terms of the order of ${\alpha}\rho/\pi$. We write it in the following form:
 \be
 K=1+K_s+K_h.
 \label{eq:eq38}
\ee
The quantity $K$ can be obtained from $\delta$, Eq.  (\ref{eq:eq31}), setting $\rho=1$. The expression of the quantity $K_h$ depends on the requested  precision. In the case when the energy of the undetected photon is small,  $\Delta\epsilon /\epsilon \ll 1$, then $K_h=0$. When it is not small or when it is not measured, then:
 \be
 K_h= \int_{\omega>\Delta\epsilon}
 \displaystyle\frac{d\sigma^{e^+e^-\to e^+e^-\gamma}}{d\sigma_0},~
d\sigma_0 =\displaystyle\frac{\pi\alpha^2}{2 s} dc \left(\displaystyle\frac{3+c^2}{1-c}\right )^2.
\label{eq:eq39}
\ee
The lower limit of the integration over $x_1$ in Eq. (\ref{eq:eq37}) is determined by the conditions:
\be
 1-\displaystyle\frac{\Delta \epsilon}{\epsilon}>x_1,~ x_2<1,~y<y_1,~z<y_2,
 \label{eq:eq310}
\ee
i.e., by the experimental conditions.
\section{Annihilation channel $e^++e^-\to 2\gamma,3\gamma$}
The cross section for electron positron annihilation in two photon   
\be
e^{-}(p_-)+e^{+}(p_+) \to \gamma(k_1)+\gamma(k_2), 
\ee
for $
s=(p_- +p_+)^2>>m_{e}^2=m^2 $
is
\be
\frac{d\sigma_{B}}{dO_1}=\frac{\alpha^2}{s\beta}\left[\frac{1+\beta^{2}c^2}{1-\beta^{2}c^2}+
2\beta^{2}(1-\beta^{2})\frac{1-c^2}{(1-\beta^{2}c^2)^2}\right],
\ee
with $\beta=\sqrt{1-4m^2/s}$, $c=\cos\theta$ ($\theta$ being the polar angle between the directions of the initial electron and of the photon in the center of mass frame). In the relativistic limit one obtains:
\be
\frac{d\sigma_{B}}{dO_1}=\frac{\alpha^{2}(1+c^2)}{s(1-c^2)}.
\ee
The theorem on factorization  of hard and soft momenta \cite{BCM83} in the expression of the cross section for exclusive processes, allows one to express the radiative corrections in the leading logarithmic approximation:
\be
\frac{\alpha}{\pi}\ll 1,\,\,\,\,\frac{\alpha}{\pi}L \sim 1,\,\,\,\,L=\log\frac{s}{m^2}
\ee
in terms of LSFs of electron and positron:
\ba
d\sigma(p_-,p_+,k_1,k_2)&=&\int\limits_0^1dx D(x,L)\cdot\int\limits_0^1dy D(y,L)d\sigma_{B}(xp_-,yp_+,k_1,k_2)
\nn\\
&&
(1+\frac{\alpha}{\pi}K_2),
\ea
where the "shifted" Born cross section has the form:
\be
d\sigma_{B}(xp_-,yp_+,k_1,k_2)=\frac{2\alpha^2}{sxy}\frac{x^2(1-c)^2+y^2(1+c)^2}
{[x(1-c)+y(1+c)]^2}dO_1.
\ee
The explicit knowledge of the $K$-factor $K=1+(\alpha/\pi)K_2$ allows one to increase the accuracy of the theoretical prediction up to $10^{-3}$ level. Such factor takes into account the non-leading contributions arising from 
the emission of virtual soft photons and hard real photons :   $K_2=K_{SV}+K_h$. The first ones was obtained many years ago \cite{B}. The result in relativistic case has the form:
\ba
K_{SV}&=&\frac{\pi^2}{3}+\frac{1}{4(1+c^2)}[(5-6c+c^2)\ln\frac{1+c}{2}+(5+2c+c^2)\ln^2\frac{1+c}{2}+ \nn \\
&&(5+6c+c^2)\ln\frac{1-c}{2}+(5-2c+c^2)\ln^2\frac{1-c}{2}].
\label{eq:eq66}
\ea
The contribution due to the emission of a hard photon close to the direction of the momentum of the electron or the positron, i.e.,  within a small cone
$\theta<\theta_0,\,\,\pi-\theta< \theta_0$,  can be obtained using the "quasi-real electrons" method.
The last part of the $K$ factor, $K_h$, is built by extracting the terms depending on $\ln \theta_0$ and adding the 
contribution form the so called non-collinear kinematical region (when the hard photon is emitted outside 
the small cones around the beams axes). The experimental 
conditions for the detection of the photons can be imposed as well. 

The cross section in non-collinear kinematics can be obtained using the chiral amplitudes method. The result is
\be
d\sigma^{hard}=\frac{16\alpha^3}{3\pi^2 s}Rd\Phi,
\ee
with
\ba
R&=&\frac{\nu_{3}^2(1+c_{3}^2)}{\nu_{1}^2\nu_{2}^2(1-c_{1}^2)(1-c_{2}^2)}+\frac{\nu_{2}^2(1+c_{2}^2)}
{\nu_{1}^2\nu_{3}^2(1-c_{1}^2)(1-c_{3}^2)}+
\nn\\
&&\frac{\nu_{1}^2(1+c_{1}^2)}{\nu_{3}^2\nu_{2}^2(1-c_{3}^2)(1-c_{2}^2)}.
\label{eq:eq67}
\ea
The phase volume element of three photon final state is
\ba
d\Phi=\frac{1}{s}\frac{d^3q_1}{2\omega_1}\frac{d^3q_2}{2\omega_2}\frac{d^3q_3}{2\omega_3}
\delta^4(p_- +p_+ -k_1-k_2-k_3).
\ea
It can be expressed in the form:
\ba
d\Phi&=&\frac{1}{16}\frac{1-\nu_1}{[2-\nu_1(1-c_{13})]^2}\nu_1d\nu_1dO_1dO_3
\nn\\
&=&
\frac{1}{16}\frac{1-\nu_3}{[2-\nu_3(1-c_{13})]^2}\nu_3d\nu_3dO_1dO_3\,\,\,...,
\ea
The final expression of $K_h$ is \cite{BB}:
\ba
\frac{\alpha}{\pi}d\sigma_{B}&(xp_-,yp_+;k_1,k_2)K_h=&\int d\sigma^{hard}\Theta+
\frac{\alpha}{2\pi}\int\limits_0^{1-\frac{\triangle E}{E}} \frac{dx}{1-x}[(1+x^2)\ln\frac{\theta_0^2}{4} \nn \\
&&+(1-x)^2]\cdot[d\sigma(xp_-,p_+)+d\sigma(p_-,xp_+)].
\ea
Here the symbol $\Theta$ represents the constraints on the manifold of integration variables $d\Phi$. The three energy fractions must exceed $\Delta E/E$ (hardness condition). Moreover, conservation laws require $\bf{k_1}+\bf{k_2}+\bf{k_3}=0$. In particular,
\be
c_1\nu_1 +c_2\nu_2 +c_3\nu_3=0, \,\,\,\nu_1 +\nu_2 +\nu_3=2,~
\nu_{i}=\frac{\omega_{i}}{E},\,\,\,\frac{\triangle E}{E}<\nu_{i}<1.
\ee
The non-collinear kinematics conditions must also be put on: $\theta_0<\theta_i<\pi-\theta_0.$
Moreover, the experimental cuts connected with detection of the final photons can be included in this
set of cuts. These conditions depend on the experimental set-up. This is the reason why we do not do numerical estimations here.
\section{Conclusions}
In case of heavy particle production ($\pi^++\pi ^-$, $\mu^++\mu ^-$), in electron--positron annihilation at energies up to several GeV, it is possible to calculate the cross section of the final particles and the interference terms  not as asymptotic quantities, as a $K$ factor, which takes into account all high order corrections. The cross section takes the form
\be
 \displaystyle\frac{d\sigma_1^{e^+e^-\to q^-q^+}}{d\Omega_-}=
 \int^1  dx_1 \int^1_2  dx_2{\cal D}(x_1,\beta){\cal D}(x_2,\beta)(1-\Pi)^{-2}
 \cdot K_i\cdot  \displaystyle\frac{d\hat \sigma_i(x_1,x_2\theta-)}{d\Omega},
\label{eq:eq42}
\ee
with $
K_i=1+\eta_i+\delta_i^{even}(\rho_e=1)+K_{ih}.$ The differential cross sections for the subprocesses
 $d\hat \sigma_{\pi,\mu}/{d\Omega_-}$ have the form:
 \ba
 \displaystyle\frac{d\hat \sigma_{\pi\pi}}{d\Omega_-}&=&
 |F_{\pi}(sx_1x_2)|^2\displaystyle\frac{\alpha^2}{2\epsilon^2}
 \displaystyle\frac{ \sin^2\theta_-}{[x_1+x_2-c_-(x_1-x_2)]^4},
 \nn\\
 \displaystyle\frac{d\hat \sigma_{\mu\mu}}{d\Omega_-}&=&
 \displaystyle\frac{\alpha^2}{2\epsilon^2}
 \displaystyle\frac{x_1^2 (1-c_-)^2+x_2^2 (1-c_-)^2}
 {[x_1+x_2-c_-(x_1-x_2)]^4}.
 \label{eq:eq43}
 \ea
In the calculation of the integral of the cross section with emission of a hard photon, difficulties appear when the photon is emitted along in the direction of the momentum of one of the charged particles ( initial or final). These so called divergences' of the integrated expression correspond to small values of the invariants which appear in the denominators.
We give a scheme, in the following calculation, which allows to overcome these 
difficulties. Let us consider the process $e^++ e^-\to\pi^++\pi ^-$. Let us choose a small angle (in radiants) $\theta_0$ such that:
\be
 \displaystyle\frac{m_e}{\epsilon} \ll\theta_0\ll 1.
 \label{eq:eq44}
\ee
Simulating the four-vector kinematics (with Monte-Carlo methods) if the angle of emission of the photon $\theta$ with respect to any of the charged particles, is large,  $ \theta>\theta_0$ it is possible to apply use the exact formulas which were given above. In such case divergences do not appear since the invariants $kp_i$ are not small 
and therefore we can neglect those terms in $\Gamma_{e^+e^-}$ which are proportional 
to $m^2_e$. 

In the case when one of the angles $\theta_i\le \theta_0$, simplified formulas can be derived, which allows a simple analytical calculation in this kinematical region \cite{Ba73}. When the photon is emitted along the directions of the beams:
$k p_i=\omega\epsilon_i(1-\beta_i\cos\theta_i)$, one has:
\be
d\sigma= d\sigma_0(p_--k,p_+)dW_{p_-}(k)+
d\sigma_0(p_-,p_+ -k)dW_{p_+}(k),
 \label{eq:eq45}
\ee
with
\be
dW_{p_{\pm}}(k)= \displaystyle\frac{\alpha}{4\pi^2}
\left [ \displaystyle\frac{1+(1-x)^2}{x\cdot k p_{\pm}}-(1-x)
\displaystyle\frac{m_e^2}{(k p_{\pm})^2} \right ]
\displaystyle\frac{d^3k}{\omega},~x=\displaystyle\frac{\omega}{\epsilon},
 \label{eq:eq46}
\ee
and $d\sigma_0$ is the Born cross section of the process
$e^+e^-\to e^{+'}e^{-'}$. In this case the three-momenta of the initial electron and 
positron are no more equal in magnitude. In the case when the electron is emitted:
$$d\sigma_0= \displaystyle\frac{2\pi\alpha^2}{ s^2}
\displaystyle\frac{s^4+t^4+u^4}{s^2t^2} dt,~s+t+u=0, $$
setting
$s=4\epsilon^2(1-x)$,
$t=- \displaystyle\frac{4(1-x)^2(1-c)}{2-x(1-c)}$ one obtains:
\be
d\sigma_0(p_--k,p_+)=\displaystyle\frac{2\pi\alpha^2}{\epsilon^2}
\left \{ \displaystyle\frac{  3-3x+x^2+2x(2-x)c+c^2(1-x+x^2)}{
(1-x)(1-c)[2-x(1-c)]^2 }\right \}^2 dc,
\label{eq:eq47}
\ee
where $c = \cos( \widehat{\vec p_-,\vec p_-^{'}})$ is the cosine of the 
scattering angle in the laboratory system.
For completeness, we introduce  the following terms for the energy and the 
angle of the scattered $e^+(e^-)$:
$$
\displaystyle\frac{\epsilon _+}{\epsilon}
=\displaystyle\frac{1}{a}[2-2x+x^2+x(2-x)c],
~\displaystyle\frac{\epsilon _-}{\epsilon}
=\displaystyle\frac{2(1-x)}{a},$$
$$
\sin\theta_+=\displaystyle\frac{\epsilon _-}{\epsilon_+}\sqrt{1-c^2},~
a=2-x(1-c).
$$
In the case when the photon is emitted from the initial positron, the cross section has the same expression under exchange of $c\to \tilde c=\cos( \widehat{\vec p_+,\vec p_+^{'}})$. In case 
of photon emission  by the scattered electrons and positrons, the cross section has the form
$c\sim \tilde c$
\be
d\sigma=d\sigma_0 [dW_{p^{'}_+}(k)+dW_{p^{'}_-}(k)],~
d\sigma_0= \displaystyle\frac{\pi\alpha^2}{8\epsilon^2}
\left(\displaystyle\frac{3+c^2}{1-c}\right )^2. 
\label{eq:eq48}
\ee
The integration of
$dW_p(k)$ over the angle, in the region $\theta <\theta_0$ gives:
\be
dW_p(k)=\displaystyle\frac{\alpha }{2\pi}\displaystyle\frac{dx }{x}
\left \{[1+(1-x)^2]\ln \left(\displaystyle\frac{\epsilon ^2\theta_0^2}{m_e^2}\right )  -(1-x)\right \}.
\label{eq:eq410}
\ee
The sum of the contributions in the regions
$\theta_i >\theta_0$ and
$\theta _i<\theta_0$
does not depend from $\theta_0$, for $\theta_0$ sufficiently small.
In the region close to the threshold of
$\pi^+\pi^-$$(\mu^+\mu^-)$, production, $\beta_{\pi}\sim\beta_{\mu}\sim 1$, 
the divergence of the cross section appears only from the radiation of the 
initial $e^+$ and $e^-$. The formula for the cross section takes the form 
(\ref{eq:eq45}), where
$s=4\epsilon^2(1-x)$, $t=m^2-2(1-x)p_-q_-$, $u=2m^2-s-t$, $x=\omega/\epsilon$:
\ba
d\sigma_0^ {e^+e^-\to \pi^+\pi^-}(p_--k,p_+)&=&\displaystyle\frac{\pi\alpha^2dt}{s^2}
\left [ 1- \displaystyle\frac{4m^2}{s} - \left ( \displaystyle\frac{t-u}{s}\right )^2\right ],
 \label{eq:eq411}
 \\
d\sigma_0 ^{e^+e^-\to \mu^+\mu^-}(p_--k,p_+)&=&\displaystyle\frac{2\pi\alpha^2dt}{s^2}
\left [  \displaystyle\frac{t^2+u^2}{s^2}+\displaystyle\frac{4m^2}{s} - 2\left ( \displaystyle\frac{m^2}{s} \right )^2\right ].
 \label{eq:eq412}
\ea
 In the ultrarelativistic limit the emission of photon along the final particles can be calculated with the help of  (\ref{eq:eq48}):
  \ba
 d\sigma^{e^+e^-\to \pi^+\pi^-}&=&\displaystyle\frac{\pi\alpha^2}{16\epsilon ^2}
 \beta^3_{\pi}\sin^2\theta dc,
 \nn\\
  d\sigma^{e^+e^-\to \mu^+\mu^-}&=&\displaystyle\frac{\pi\alpha^2}{8\epsilon ^2}
 \beta_{\mu}(2-\beta_{\mu}^2\sin^2\theta) dc.
 \label{eq:eq413}
\ea
In conclusion, we have calculated the cross section for different processes induced by $e^+e^-$ annihilation, in the energy range 200 MeV $\le 2E\le$ 3 GeV, where the contribution of heavy bosons can be safely neglected. We have taken into account first order corrections to the amplitudes and the corrections
due to soft emitted photons. The charge--odd and charge--even contributions to the cross section for the final channels $\pi^+\pi^-(\gamma)$ and $\mu^+\mu^-(\gamma)$ have been explicitly given. In case of pions,
form factors are taken into account. The comparison with the lepton structure function method allows to estimate the contribution of
higher orders of perturbation theory. The precision of the obtained results is better than 0.5\%, in the energy region outside narrow resonances.
Finally we have described a method to integrate the cross section, avoiding the difficulties which may arise from singularities in specific kinematical regions.


\end{document}